\documentclass[twocolumn,prd,superscriptaddress]{revtex4-1}
\usepackage[dvipsnames]{xcolor} 
\usepackage{amsmath} 
\usepackage{booktabs}
\usepackage[utf8]{inputenc}
\usepackage{graphicx} 
\usepackage[colorlinks=true,citecolor=blue,urlcolor=blue]{hyperref}
\usepackage[normalem]{ulem}
\usepackage{xspace}

\newcommand{\SSM}{SSM\xspace}
\newcommand{\OhThreeAStart}{1 April 2019 1500 UTC}
\newcommand{\OhThreeAEnd}{1 October 2019 1500 UTC}

\newcommand{\SNRLossFromFlow}{9\%}
\newcommand{\VolLossFromFlow}{24\%\xspace}
\newcommand{\Flow}{45 Hz}
\newcommand{\msun}{\ensuremath{M_\odot}\xspace}
\newcommand{\MinMatch}{\ensuremath{0.97}\xspace}

\newcommand{\MinPrimaryMass}{\ensuremath{0.2}\xspace}
\newcommand{\MinSecondaryMass}{\ensuremath{0.2}\xspace}
\newcommand{\MaxPrimaryMass}{\ensuremath{10 }\xspace}
\newcommand{\MaxSecondaryMass}{\ensuremath{1.0}\xspace}
\newcommand{\MaxLowMassComponentSpin}{\ensuremath{0.1}\xspace}
\newcommand{\MaxHighMassComponentSpin}{\ensuremath{0.9}\xspace}
\newcommand{\MinMassRatio}{\ensuremath{0.1}\xspace}

\newcommand{\PBHUpperVague}{6\%\xspace}

\newcommand{\fdbhdef}{\ensuremath{f_\mathrm{DBH}\equiv \Omega_{\mathrm{DBH}}/\Omega_\mathrm{DM}}~\xspace}
\newcommand{\fpbhdef}{\ensuremath{f_\mathrm{PBH}\equiv \Omega_{\mathrm{PBH}}/\Omega_\mathrm{DM}}}
\newcommand{\fdbh}{\ensuremath{f_\mathrm{DBH}}~\xspace}
\newcommand{\fpbh}{\ensuremath{f_\mathrm{PBH}}~\xspace}
\newcommand{\mmin}{\ensuremath{M_{\rm min}}\xspace}
\newcommand{\DBHatOneMsun}{\ensuremath{0.003\%}\xspace}
\newcommand{\MinFarCandidate}{0.14 $\mathrm{yr}^{-1}$}
\newcommand{\RatesWorst}{\ensuremath{ 24200 }\xspace}
\newcommand{\RatesBest}{\ensuremath{ 220 }\xspace}
\newcommand{\ratesunits}{\ensuremath{\mathrm{Gpc}^{-3} \, \mathrm{yr}^{-1}}~\xspace}

\newcommand{\MaxRedshift}{0.12\xspace}

\newcommand{\CalibrationUncertainty}{$\sim 3\%$\xspace}
\newcommand{\RatesCalibrationUncertainty}{$\sim10\%$\xspace}

\newcommand{\LVK}{LVK\xspace}
\newcommand{\OhThreeA}{O3a\xspace}
\newcommand{\paper}{{\it Letter}\xspace}

\def\prd{{\it Physical Review D}\xspace}

\def\apj{{\it Astrophysical Journal}\xspace}

\def\mnras{{\it Monthly Notices of the RAS}\xspace}

\begin{document}

\title{Search for subsolar-mass binaries in the first half of Advanced LIGO and Virgo's third observing run}



\author{R.~Abbott}
\affiliation{LIGO Laboratory, California Institute of Technology, Pasadena, CA 91125, USA}
\author{T.~D.~Abbott}
\affiliation{Louisiana State University, Baton Rouge, LA 70803, USA}
\author{F.~Acernese}
\affiliation{Dipartimento di Farmacia, Universit\`a di Salerno, I-84084 Fisciano, Salerno, Italy}
\affiliation{INFN, Sezione di Napoli, Complesso Universitario di Monte S. Angelo, I-80126 Napoli, Italy}
\author{K.~Ackley}
\affiliation{OzGrav, School of Physics \& Astronomy, Monash University, Clayton 3800, Victoria, Australia}
\author{C.~Adams}
\affiliation{LIGO Livingston Observatory, Livingston, LA 70754, USA}
\author{N.~Adhikari}
\affiliation{University of Wisconsin-Milwaukee, Milwaukee, WI 53201, USA}
\author{R.~X.~Adhikari}
\affiliation{LIGO Laboratory, California Institute of Technology, Pasadena, CA 91125, USA}
\author{V.~B.~Adya}
\affiliation{OzGrav, Australian National University, Canberra, Australian Capital Territory 0200, Australia}
\author{C.~Affeldt}
\affiliation{Max Planck Institute for Gravitational Physics (Albert Einstein Institute), D-30167 Hannover, Germany}
\affiliation{Leibniz Universit\"at Hannover, D-30167 Hannover, Germany}
\author{D.~Agarwal}
\affiliation{Inter-University Centre for Astronomy and Astrophysics, Pune 411007, India}
% ...

\collaboration{The LIGO Scientific Collaboration, the Virgo Collaboration, and the KAGRA Collaboration}



\author{Donghui Jeong}
\affiliation{The Pennsylvania State University, University Park, PA 16802, USA}
\author{Sarah Shandera}
\affiliation{The Pennsylvania State University, University Park, PA 16802, USA}

\begin{abstract}

We report on a search for compact binary coalescences where at least one
binary component has a mass between \MinSecondaryMass~\msun and \MaxSecondaryMass~\msun in Advanced LIGO and
Advanced Virgo data collected between \OhThreeAStart{} and \OhThreeAEnd{}. 
We extend previous analyses in two main ways: we include data from the Virgo detector and we allow for more unequal mass systems, with mass ratio $q \geq \MinMassRatio$.
We do not report any gravitational-wave candidates. The most significant trigger has a false alarm rate of \MinFarCandidate. This implies an upper limit on the merger rate of subsolar binaries in the range $[\RatesBest \textendash \RatesWorst]~\ratesunits$, depending on the chirp mass of the binary.  We use this upper limit to derive astrophysical constraints on two phenomenological models that could produce subsolar-mass compact objects. One is {an isotropic distribution of equal-mass} primordial black holes. Using this model, we find that the fraction of dark matter in primordial black holes is $\fpbhdef\lesssim \PBHUpperVague$. The other is a dissipative dark matter model, in which fermionic dark matter can collapse and form black holes. The upper limit on the fraction of dark matter black holes depends on the minimum mass of the black holes that can be formed: the most constraining result is obtained at $\mmin=1~\msun$, where $\fdbhdef\lesssim\DBHatOneMsun$. These are the tightest limits on spinning subsolar-mass binaries to date.

\end{abstract}

\maketitle

\section{Introduction}\label{sec:intro}

The first detection of gravitational waves from a binary black hole (BBH) merger in 2015~\cite{Abbott:2016blz} has given us a new way to study the universe. Since then, dozens of gravitational waves (GWs) have been detected in Advanced
LIGO~\cite{TheLIGOScientific:2014jea} and Advanced
Virgo~\cite{TheVirgo:2014hva} data.
The LIGO Scientific, Virgo, and KAGRA Collaboration (\LVK) have reported the discovery of GWs from approximately fifty binary black holes (BBHs), binary neutron stars (BNSs), or neutron star black hole
mergers (NSBHs)~\cite{LIGOScientific:2018mvr,Abbott:2020niy,LIGOScientific:2021qlt}.
Further analyses on public data~\cite{Abbott:2019ebz,Trovato:2019liz} have resulted in the discovery of other compact binaries~\cite{Zackay:2019tzo,Venumadhav:2019tad,Venumadhav:2019lyq,Nitz:2019hdf,Zackay:2019btq}.
The gravitational-wave sources presented in~\cite{LIGOScientific:2018mvr,Abbott:2020niy} are already being used to answer key questions including cosmological measurements~\cite{Abbott:2017xzu,Abbott:2019yzh,Soares-Santos:2019irc,Vasylyev:2020hgb,Finke:2021aom}, analyses of the mass and spin distribution of compact objects, their formation channels~\cite{Abbott:2020gyp,Roulet:2018jbe,Roulet:2021hcu,Callister:2021fpo,Wong:2020ise,Kimball:2020qyd,Zevin:2020gbd,Bouffanais:2021wcr,Franciolini:2021tla}, and tests of general relativity~\cite{Abbott:2020jks,Carullo:2021yxh,Isi:2019aib}

The black holes detected with gravitational waves can have masses larger than those discovered in X-ray binaries~\cite{Ozel:2010su,Farr:2010tu,Jonker:2021rkb,Miller-Jones:2021plh}. Several GW sources have challenged our understanding of astrophysics and stellar evolution~\cite{Abbott:2020uma,Abbott:2020tfl,Abbott:2020mjq,Abbott:2020khf,Gerosa:2020bjb,Rodriguez:2020viw,Hamers:2020huo,Kimball:2020opk,Liu:2020gif}. One such source is GW190521~\cite{Abbott:2020tfl,Abbott:2020mjq}, a system whose most massive black hole might have a mass in the pair instability mass gap~\cite{Woosley:2016hmi,Abbott:2020mjq,Costa:2020xbc,Baxter:2021swn} (but see e.g. Refs~\cite{Kremer:2020wtp,Renzo:2020smh,Farrell:2020zju,Nitz:2020mga,Belczynski:2020bca,Sakstein:2020axg}). With a mass of $\sim 142$~\msun, the merger product of GW190521 was likely an intermediate mass black hole~\cite{Abbott:2020mjq,Greene:2019vlv}. At the other end of the mass spectrum, the lightest object in GW190814, a $\sim 2.6$~\msun compact object, was either the heaviest neutron star or the lightest black hole ever discovered~\cite{Abbott:2020khf,Most:2020bba,Tews:2020ylw,Dexheimer:2020rlp,Nathanail:2021tay}. 

There are no widely accepted astrophysical channels that predict the formation of subsolar-mass (\SSM) objects significantly more compact than white dwarfs.
Since the endpoint of stellar evolution of massive stars is either a neutron star or a supersolar-mass black hole, the existence of a compact object below 1 \msun would be indicative of a new formation mechanism, and potentially of new physics. 

One possible scenario for the formation of \SSM black holes is the collapse of overdensities in the early universe, resulting in primordial black holes (PBHs)~\cite{Hawking:1971ei,Zeldovich1967,1975ApJ...201....1C,1975Natur.253..251C}. The amplitude of primordial fluctuations on very small scales \cite{GarciaBellido:1996qt,Clesse:2015wea}, together with the equation-of-state of the early universe \cite{Jedamzik:1996mr, Jedamzik:1998hc}, determines the mass and abundance of these objects \cite{Byrnes:2018clq,Carr:2019kxo}. In particular, their masses might be in the range probed by ground-based detectors~\cite{Clesse:2015wea,Bird:2016dcv,Sasaki:2016jop}, and so the mass spectrum is constrained by gravitational-wave data~\cite{Clesse:2017bsw,Diego:2019rzc,Gow:2019pok,DeLuca:2020qqa,Hall:2020daa,Jedamzik:2020omx,Clesse:2020ghq}. 
Alternatively, if dark matter has a sufficiently complex particle composition, which allows for chemistry and dissipation, small compact objects could form through the cooling and gravitational collapse of dark matter halos~\cite{DAmico:2017lqj,Shandera:2018xkn,Choquette:2018lvq}. 
If dark matter is sufficiently dissipative, compact objects would form through pathways similar to known astrophysical channels, with details dependent on the interactions specific to the dark
sector. Dissipative dark matter models that produce black holes in the subsolar to supersolar range were recently constrained in~\cite{Singh:2020wiq} by analyzing \LVK data.
Another possibility is that ultralight bosonic fields clump together to form self-gravitating, horizonless compact objects, known as boson stars \cite{Kaup:1968zz, Ruffini:1969qy, Brito:2015pxa}. Their maximum mass depends on the mass of the bosonic particle, hence they might be subsolar if the latter is larger than $10^{-10}$~ eV/$c^2$~\cite{Liebling:2012fv, Herdeiro:2017fhv}.
Finally, some dark matter models predict the formation of $\sim 1 M_\odot$ black holes through the accumulation of dark matter particles in neutron star cores~\cite{Kouvaris:2010jy, deLavallaz:2010wp,PhysRevD.40.3221, Bramante:2015dfa,
Bramante:2014zca,Bramante:2017ulk,Kouvaris:2018wnh}. Black holes formed via this class of mechanisms would have masses comparable to or smaller than the mass of the neutron star.

Searches for compact binaries with at least one \SSM component have
been carried out in both Initial LIGO~\cite{Abbott:2005pf,Abbott:2007xi,wadethesis} and Advanced LIGO and Advanced Virgo data~\cite{Abbott:2018oah,Authors:2019qbw}. Advanced LIGO and Advanced Virgo data have more recently been analyzed in~\cite{Nitz:2020bdb,Nitz:2021mzz,Phukon:2021cus} for systems with lower mass ratios and higher eccentricities than those considered by the \LVK. No detections were reported.  In this \paper, we report the results of searches for \SSM{} compact
binaries in the first half of Advanced LIGO and Advanced Virgo's third observing run (this is the first half of the third science run, henceforth \OhThreeA). While no sources are detected, we obtain limits on the abundance of monochromatic PBHs and black holes formed by dissipative fermionic dark matter.

\section{Search}\label{sec:search}

The data used for this \paper were collected during \OhThreeA by the Advanced LIGO and
Advanced Virgo interferometers between \OhThreeAStart{} and \OhThreeAEnd{}. The data characterization and calibration were performed as described in Refs.~\cite{Sun:2020wke,Acernese:2018bfl,Abbott:2020niy,Davis:2021ecd} with the addition of a non-linear removal of spectral lines~\cite{Vajente:2019ycy,o3afinal}.

We present results from three matched-filter based pipelines:
GstLAL~\cite{Messick:2016aqy,Sachdev:2019vvd,Hanna:2019ezx},
MBTA~\cite{Aubin:2020goo}, and
PyCBC~\cite{Allen:2005fk,Allen:2004gu,Canton:2014ena,Usman:2015kfa,Nitz:2017svb,Davies:2020tsx}. These analyses correlate the data with a bank of templates that model the gravitational-wave signals expected from binaries in quasi-circular orbit.
The bank is designed to recover binaries with (redshifted) primary mass $m_1 \in$ [\MinPrimaryMass, \MaxPrimaryMass]\msun and  secondary mass $m_2 \in$[\MinSecondaryMass, \MaxSecondaryMass]\msun . We additionally limit the binary mass ratio, $q \equiv m_2 / m_1$, to range from $\MinMassRatio < q < 1.0$. We include the effect of spins aligned with the orbital angular momentum in the gravitational waveform used in the template bank~\cite{Brown:2012qf}. 
When a binary component, $m_i$, has a mass $m_i \geq 0.5 \thinspace M_\odot$, 
we allow for a dimensionless component spin up to \MaxHighMassComponentSpin{}.
For compact objects with $m_i < 0.5 \thinspace M_\odot$, we limit the maximum dimensionless
spin to \MaxLowMassComponentSpin{}. We chose to restrict the possible spin magnitude in the low-mass part of the template bank, and not to allow for spin precession in order to reduce the computational cost.
All three searches use the same template bank, constructed using a geometric placement algorithm~\cite{Harry:2013tca} with a minimum
match~\cite{Owen:1995tm} of \MinMatch. This ensures that no more than 10\% of astrophysical signals can be missed due to the discrete template placement. 
We use the \texttt{TaylorF2} waveform~\cite{Sathyaprakash:1991mt, Blanchet:1995ez, Poisson:1997ha, Damour:2001bu, Mikoczi:2005dn, Blanchet:2005tk, Arun:2008kb,Buonanno:2009zt,Bohe:2013cla,Bohe:2015ana,Mishra:2016whh}, including phase terms {up} to 3.5 post-Newtonian order, but no amplitude corrections.

This search covers a larger mass and spin range than the last \LVK analysis for \SSM{} objects~\cite{Authors:2019qbw}. 
As a result, we require 
approximately twice as many template waveforms to effectively cover the search parameter space. To
reduce the computational cost of the search, we analyze the data from \Flow{} instead of 15 Hz (as in the searches described in~\cite{Abbott:2020niy}). We estimate that this restricted bandwidth results in a maximum loss of signal-to-noise ratio (SNR) of \SNRLossFromFlow{}, relative to what would be obtained filtering from 15~Hz. In turn, this results in a maximum reduction of the surveyed volume of \VolLossFromFlow.

The three pipelines used in this \paper are described in more detail in Ref.~\cite{Abbott:2020niy,o3afinal}. Here we only highlight differences in the way each pipeline has been run for this analysis, as compared to Ref.~\cite{Abbott:2020niy,o3afinal}.

GstLAL's~\cite{Messick:2016aqy,Sachdev:2019vvd,Hanna:2019ezx} detection statistic is unchanged relative to Ref.~\cite{Abbott:2020niy}. GstLAL reweighs waveforms in the template bank according to the characteristics of the expected population~\cite{Fong:2018elx}. However, because \SSM{} populations are yet to be observed we use a population model uniform in template density for this search. GstLAL uses a similar procedure to the one it employed in Ref.~\cite{Abbott:2020khf} and includes all events from the analyzed period in the noise background to provide a conservative false-alarm-rate estimate. 
As in previous \SSM{} searches~\cite{Abbott:2018oah,Authors:2019qbw} we do not use a gating scheme to account for loud noise artifacts~\cite{Messick:2016aqy}; instead we rely on statistical data quality information from the  iDQ algorithm~\cite{Essick:2020qpo,Godwin:2020weu}.

The MBTA pipeline splits the matched filtering in two different frequency bands in order to reduce the computational cost~\cite{Babak:2008rb,Privitera:2013xza}. 
The set-up of the search is unchanged with respect to Ref.~\cite{Aubin:2020goo} with two exceptions in order to adapt to the extended duration of low mass binaries: we use longer stretches of data to perform fast Fourier transforms (FFTs) and to calculate the noise power spectral density (PSD).
For the FFT, we use from seconds to hundreds of seconds of data, while the PSD update time is up to two times longer than for standard BNS searches, depending on the frequency region under consideration.

The PyCBC pipeline~\cite{Allen:2005fk, Canton:2014ena, Usman:2015kfa, Nitz:2017svb, Davies:2020tsx, pycbcsoft} 
is unchanged relative to the configuration described in Ref.~\cite{o3afinal}.
However, the sine-Gaussian veto described in Ref.~\cite{Nitz:2017lco} is not used, due to the low total mass of the template bank.

\section{Results}\label{sec:results}

No gravitational-wave candidates were identified by any of the search pipelines. The most significant candidate has a false-alarm rate (FAR) of \MinFarCandidate{}. The lack of detections can be recast as an upper limit on the merger rate of compact binaries. 
First, we estimate the sensitivity of each search pipeline for binaries in a given population. This can be done by computing the surveyed time-volume:

\begin{equation} \langle VT \rangle = T\int \mathrm{d}z \frac{\mathrm{d}V}{(1+z) \mathrm{d}z} \epsilon(z);, \label{Eq.VTdef}
\end{equation}
where $T$ is the analyzed time and $\epsilon(z)$ is the efficiency. The efficiency represents the fraction of astrophysical sources in the population which are detectable at a redshift $z$. The efficiency can be written as the probability that a binary with parameters $\vec{\theta}$ is detectable (a quantity often referred to as $p_{\rm{det}}(\vec\theta)$ in the literature, e.g. Ref.~\cite{Abbott:2020gyp}) integrated over the distribution of all parameters but the redshift. Therefore, in order to calculate Eq.~(\ref{Eq.VTdef}) we need to assume a model for the mass distribution, spin distribution, sky positions and orbital orientations~\cite{Mandel:2018mve,Thrane:2018qnx,Vitale:2020aaz}. Since we are only sensitive to nearby ($z\lesssim \MaxRedshift$) sources we treat the merger rate as constant.

Each pipeline estimates its sensitivity by simulating
gravitational-wave signals from a population of \SSM compact binaries and adding them into the collected data. We simulate a population with a uniform distribution of source masses in the range $0.2 M_\odot < m_1 < 10.0 M_\odot$ and $0.2 M_\odot < m_2 < 1.0 M_\odot$. We make an additional detector frame mass cut $m_1 < 10 M_\odot$ ($m_2 < 1 M_\odot$) due to the template bank coverage. We reject binaries with mass ratios exceeding the $0.1 < q < 1.0$ bounds of our search. The dimensionless spins are again assumed to be oriented in the direction of the angular momentum for computational reasons. The spin magnitude is uniform  in the range $-0.1 < \chi_{k} < 0.1$ ($-0.9 < \chi_{k} < 0.9$) when $m_k < 0.5 M_\odot$ ($m_k > 0.5 M_\odot$). The sources are uniform in comoving volume, isotropically distributed on the sphere, and randomly oriented. We use the Planck “TT,TE,EE+lowP+lensing+ext” cosmology~\cite{Ade:2015xua}.

Since the search sensitivity is primarily a function of chirp mass, $\mathcal{M} \equiv (m_1 m_2)^{3/5} / (m_1 + m_2)^{1/5}$~\cite{Cutler:1994ys}, we break this population into 9 equally spaced chirp mass bins {in the range $0.17 M_\odot < \mathcal{M} < 2.39 M_\odot$} to determine $\langle VT \rangle(\mathcal{M})$.

\begin{figure}
\includegraphics[width=\columnwidth]{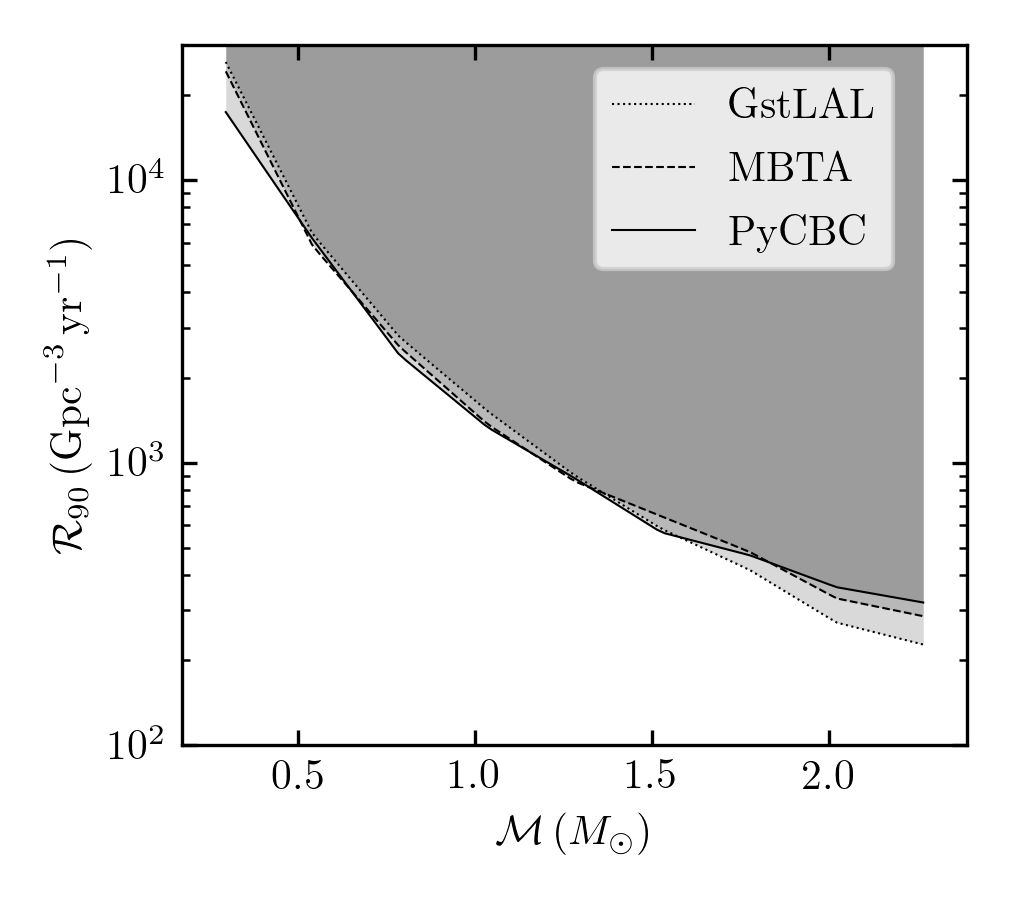}
\caption{\label{fig:Rates} Upper limit on the merger rate of binaries with at least one \SSM{} component as a function of source frame chirp mass. The dotted, dashed, and solid lines represent the 90\% confidence limits obtained by GstLAL, MBTA, and PyCBC, respectively.
}
\end{figure}

Treating each chirp mass bin as a different population, labeled by an index $i$, we can use the surveyed time-volume~\cite{Tiwari:2017ndi} ${\langle VT \rangle_i}$ for each chirp mass bin to estimate a frequentist upper limit (90\% confidence interval) on the merger rate of that population by using the loudest event statistic~\cite{Abbott:2018oah,Authors:2019qbw,Biswas:2007ni}:

\begin{equation} \mathcal{R}_{90,i}= \frac{2.3}{\langle VT \rangle_i} .\end{equation}
This is shown in Fig.~\ref{fig:Rates} for the three pipelines. Although the pipelines generally agree, differences in background estimation and ranking statistics can lead to $\langle VT \rangle$ measurements that agree to within $\mathcal{O}(30\%)$. In what follows, we use the MBTA results as our fiducial rate constraint.
Instrumental calibration errors were at most \CalibrationUncertainty in amplitude in the bandwidth relevant for our analysis, and usually much smaller~\cite{Sun:2020wke}. At most, they could contribute a \RatesCalibrationUncertainty uncertainty in our $\langle VT \rangle_i$ measurement. We follow~\cite{o3afinal,Abbott:2020gyp} and neglect their impact in the remainder of this work.

For any astrophysical model that could generate \SSM{} binaries, the merger rate upper limits can be used to set constraints on the model parameters. Here we focus on two such models: formation of PBHs catalyzed by three-body interactions~\cite{Ioka:1998nz}, and dark-matter black holes formed by late-time gravitational collapse of dark matter sub-structure \cite{Shandera:2018xkn}. 

\begin{figure}[th]
\includegraphics[width=\columnwidth]{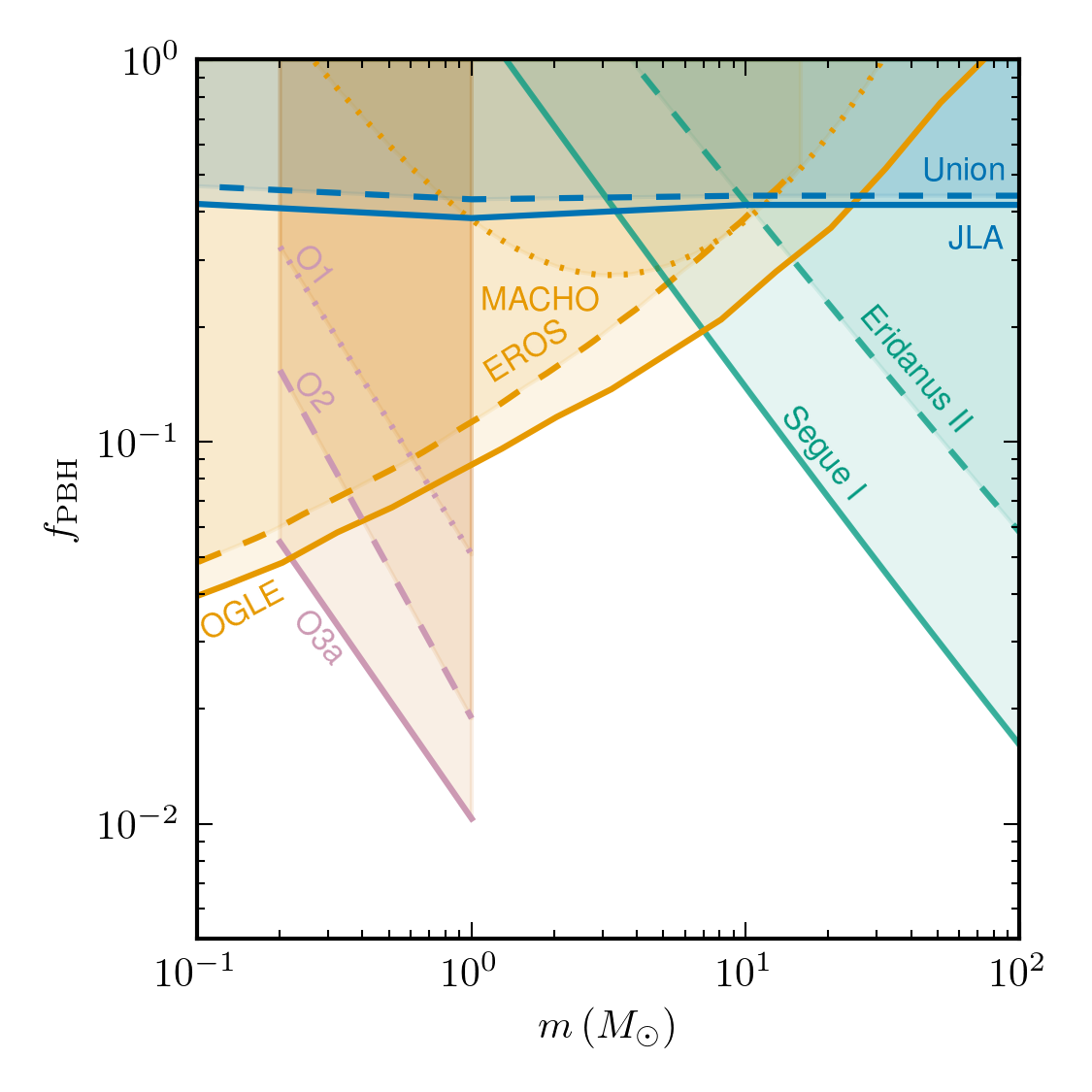}
\caption{\label{fig:upperlimits} Constraints on the fraction of dark matter in PBHs. The horizontal axis shows the source frame mass of the black hole in each model; for \LVK results this is the component mass for each object in the binary. Each constraint shown carries a model dependency. Shown (pink) are the \LVK results from O1~\cite{Abbott:2018oah}, O2~\cite{Authors:2019qbw}, and \OhThreeA (this work); (orange) microlensing constraints from MACHO~\cite{Allsman:2000kg}, EROS~\cite{Tisserand:2006zx}, and OGLE~\cite{2011MNRAS.416.2949W}; (green) dynamical constraints from observations of Segue I~\cite{Koushiappas:2017chw} and Eridanus II~\cite{Brandt:2016aco} dwarf galaxies; (blue) supernova lensing constraints from the Joint Light-curve Analysis and Union 2.1 datasets~\cite{Zumalacarregui:2017qqd}. \LVK results use the Planck
“TT,TE,EE+lowP+lensing+ext” cosmology~\cite{Ade:2015xua}.}
\end{figure}

We use a phenomenological model for PBHs, rather than a first-principles model derived from an inflationary potential (see for example ~\cite{Byrnes:2018txb,Kalaja:2019uju} for work connecting PBH distributions to inflationary models). Following ~\cite{Ioka:1998nz} we assume PBHs produced at a single mass, and randomly distributed in space (see Appendix \ref{PBHAppendix} for details). This model predicts a merger rate given the mass of the PBHs in the binary and the abundance of PBHs, parametrized as a fraction of the dark matter density, \fpbhdef{}. By using the merger rate upper limits derived above, we can thus obtain an upper limit on \fpbh as a function of the component mass of the black holes in the binary~\cite{Ioka:1998nz}. This is shown in Fig.~\ref{fig:upperlimits}.

In this analysis, it is assumed that the two objects in the binary have the same mass. Because the detectors' sensitivity depends more strongly on the chirp mass than on the mass ratio, for this analysis we assume that the rate upper limits we presented above (which included unequal mass binaries) can be used to assess the rate of equal mass binaries: $\mathcal{R}_{90}(\mathcal{M}, q=1 ) \approx \mathcal{R}_{90}(\mathcal{M})$.  
Under these assumptions, we find $\fpbh \lesssim \PBHUpperVague$ for PBHs in equal-mass binaries with component objects in the range $[\MinSecondaryMass -\MaxSecondaryMass]~\msun$. The method of Ref.~\cite{Bellomo:2017zsr} may be used to interpret these constraints on generic PBH mass functions.
Recent work \cite{Ali-Haimoud:2017rtz, Kavanagh:2018ggo} has shown that there are a number of mechanisms that can alter and suppress the PBH merger rate from that derived in Ref.~\cite{Ioka:1998nz} and used here; these include binary disruption from other close PBHs, clusters of black holes, and matter inhomogeneities~\cite{Raidal:2018bbj}. Suppression of the theoretical merger rate leads to looser constraints on the allowable fraction of the dark matter contained in PBHs. 

Next, we consider a dissipative dark-matter model which consists of two fermions, oppositely charged under a dark version of electromagnetism, together with a massless dark photon. The dark matter can form bound states analogous to atomic and molecular hydrogen, and dissipate energy by radiative processes including Bremsstrahlung, recombination, and collisional excitation \cite{Buckley:2017ttd}. In dense regions, some dark matter gas can cool efficiently enough for gravitational collapse to proceed, eventually forming black holes~\cite{Shandera:2018xkn}. In contrast to the PBH case, here we assume a power-law distribution for the black hole masses, with an unknown low-mass cutoff.
We calculate an upper limit on the fraction of the dissipative dark matter that ends up in black holes ($\fdbhdef$) as a function of the low-mass cutoff for the dark matter black holes, marginalized over all other parameters of the model (e.g. the slope of the dark matter black hole mass function). More details on the model are given in Appendix~\ref{App.DBHs}.
In Fig.~\ref{fig:dissipativeDM}, we show our constraints. The lowest upper limit is found at $\mmin=1~\msun$, where $\fdbh \lesssim \DBHatOneMsun$. No meaningful constraints can be set for $\mmin\lesssim 2\times10^{-2}~\msun$ since below that mass none of the black holes in the population would be detectable with the current sensitivity, hence a non-detection does not yield any constraint. 

\begin{figure}[th]
\includegraphics[width=\columnwidth]{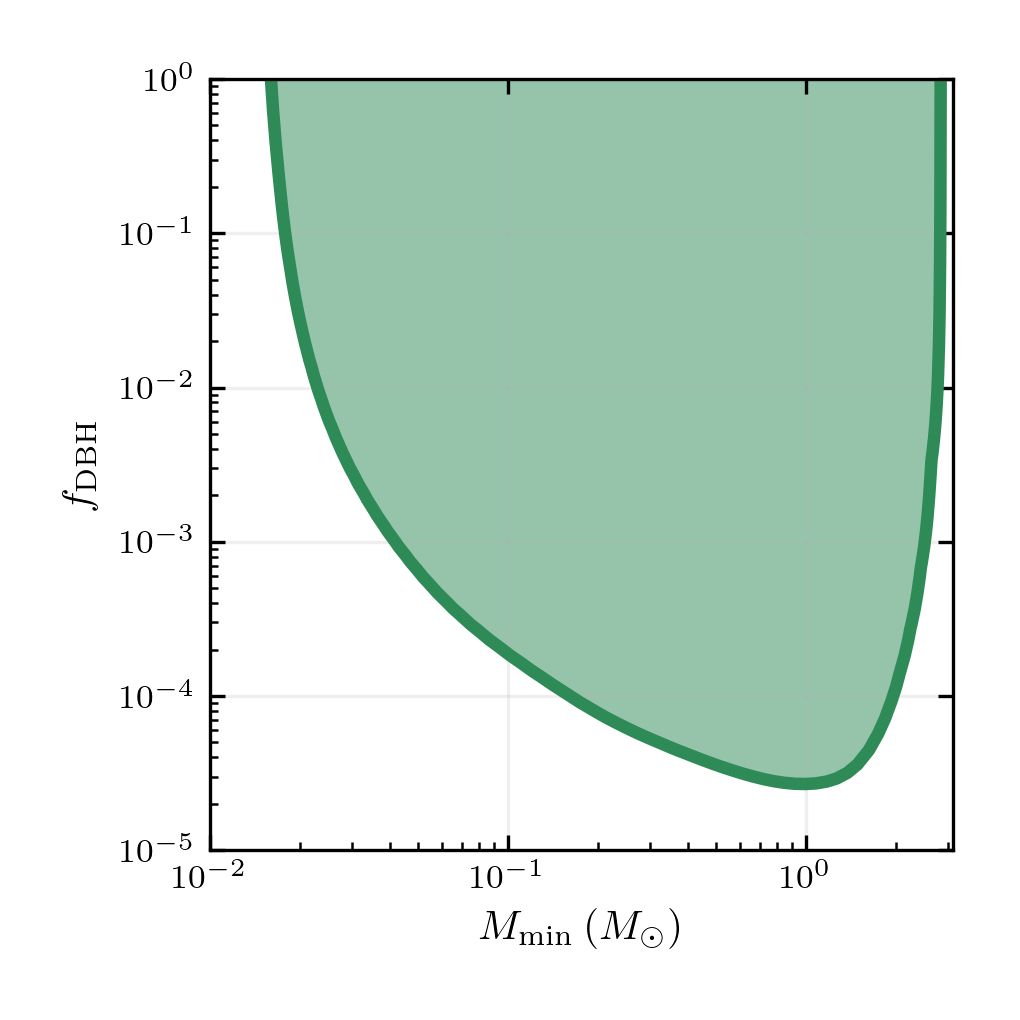}
\caption{\label{fig:dissipativeDM} Constraints on the fraction of dark matter, $f_\mathrm{DBH}$, in black holes formed from cooling of dissipative dark matter and their minimum possible source frame mass $M_{\rm min}$. 
}
\end{figure}

\section{Conclusions and outlook}\label{sec:conc}

Gravitational waves from compact object mergers provide a unique probe of dark matter structures on the smallest scales. Here, we have considered two possible dark matter candidates: PBHs and fermionic dark matter particles that can dissipate and form dark matter black holes. Both of these formation mechanisms can potentially produce both sub and supersolar mass black holes. We have focused on the \SSM{} regime, which cannot be populated with black holes by any known astrophysical channel. 

We have used three different algorithms to search the data from O3a for compact binaries in which at least one of the component objects had a mass between $[0.2-1.0]~\msun$.
We have found no candidates, and obtained upper limits on the merger rate of \SSM{}  black holes in the range $[\RatesBest-\RatesWorst]~\ratesunits$. The upper limit is dependent on the chirp mass of the source and shown in Fig.~\ref{fig:Rates}. These upper limits can be recast into limits on the physical parameters of \SSM{} black holes populations. 

By considering a phenomenological model for \SSM{} PBHs in which the compact objects are all formed with the same mass, we have obtained a limit on the abundance of these black holes as a function of their mass at formation: $\fpbh \lesssim \PBHUpperVague$ in the mass range, as seen in Fig.~\ref{fig:upperlimits}. This significantly improves microlensing and supernova lensing constraints in the same mass region as well as our previous constraints from Ref.~\cite{Authors:2019qbw}, though we note that there are uncertain mechanisms that can reduce the expected PBH merger rate and raise the allowed value of $f_\mathrm{PBH}$~\cite{Ali-Haimoud:2017rtz, Kavanagh:2018ggo,Raidal:2018bbj}.
We have also considered a model for fermionic dissipative dark matter, parametrized by the abundance of the black holes it produces, and by their minimum mass. 
The most stringent limit is obtained at $\mmin=1~\msun$ for which $\fdbh \lesssim \DBHatOneMsun$, as shown in Fig.~\ref{fig:dissipativeDM}.
 The constraint on the minimum mass can be interpreted in two ways. The most straightforward is as a constraint on the Chandrasekhar limit of dark matter black holes~\cite{Shandera:2018xkn}, which constrains the mass of a dark fermion analogous to the proton to be in the range $0.66 \textendash 8.8$ GeV/$c^2$. Additionally, the minimum mass of black holes formed when the dark matter gas cools and fragments depends on the coldest temperature the gas can reach, that is, on the dark matter chemistry. For the model we considered, this temperature is set by the energy difference of the lowest energy molecular radiative transition. Therefore, a constraint on the minimum mass of any dark black holes also implies a constraint on the dark molecular energy spacing, although the precise relationship depends on astrophysical modeling. 

In the coming years, the sensitivity of Advanced LIGO and Advanced Virgo will continue to improve~\cite{Aasi:2013wya}, and the global network of detectors is expected to grow with the addition of KAGRA~\cite{Akutsu:2020his} and LIGO-Aundha~\cite{M1100296}. These advances will allow for more stringent limits in the near future, or even the detection of a \SSM compact object.

\emph{Note} As our work was finalized, Ref.~\cite{Nitz:2021vqh} reported results on a search for binaries with no spin and component masses $m_1 \in (0.1 M_\odot, 7.0 M_\odot), m_2 \in (0.1 M_\odot, 1.0 M_\odot)$ in O3a data. That search also reported no detections.  

\section*{Acknowledgments}

This material is based upon work supported by NSF's LIGO Laboratory which is a major facility
fully funded by the National Science Foundation.
The authors also gratefully acknowledge the support of
the Science and Technology Facilities Council (STFC) of the
United Kingdom, the Max-Planck-Society (MPS), and the State of
Niedersachsen/Germany for support of the construction of Advanced LIGO
and construction and operation of the GEO600 detector.
Additional support for Advanced LIGO was provided by the Australian Research Council.
The authors gratefully acknowledge the Italian Istituto Nazionale di Fisica Nucleare (INFN),
the French Centre National de la Recherche Scientifique (CNRS) and
the Netherlands Organization for Scientific Research (NWO),
for the construction and operation of the Virgo detector
and the creation and support  of the EGO consortium.
The authors also gratefully acknowledge research support from these agencies as well as by
the Council of Scientific and Industrial Research of India,
the Department of Science and Technology, India,
the Science \& Engineering Research Board (SERB), India,
the Ministry of Human Resource Development, India,
the Spanish Agencia Estatal de Investigaci\'on,
the Vicepresid\`encia i Conselleria d'Innovaci\'o, Recerca i Turisme and the Conselleria d'Educaci\'o i Universitat del Govern de les Illes Balears,
the Conselleria d'Innovaci\'o, Universitats, Ci\`encia i Societat Digital de la Generalitat Valenciana and
the CERCA Programme Generalitat de Catalunya, Spain,
the National Science Centre of Poland and the European Union - European Regional Development Fund; Foundation for Polish Science (FNP),
the Swiss National Science Foundation (SNSF),
the Russian Foundation for Basic Research,
the Russian Science Foundation,
the European Commission,
the European Regional Development Funds (ERDF),
the Royal Society,
the Scottish Funding Council,
the Scottish Universities Physics Alliance,
the Hungarian Scientific Research Fund (OTKA),
the French Lyon Institute of Origins (LIO),
the Belgian Fonds de la Recherche Scientifique (FRS-FNRS),
Actions de Recherche Concertées (ARC) and
Fonds Wetenschappelijk Onderzoek - Vlaanderen (FWO), Belgium,
the Paris \^{I}le-de-France Region,
the National Research, Development and Innovation Office Hungary (NKFIH),
the National Research Foundation of Korea,
the Natural Science and Engineering Research Council Canada,
Canadian Foundation for Innovation (CFI),
the Brazilian Ministry of Science, Technology, and Innovations,
the International Center for Theoretical Physics South American Institute for Fundamental Research (ICTP-SAIFR),
the Research Grants Council of Hong Kong,
the National Natural Science Foundation of China (NSFC),
the Leverhulme Trust,
the Research Corporation,
the Ministry of Science and Technology (MOST), Taiwan,
the United States Department of Energy,
and
the Kavli Foundation.
The authors gratefully acknowledge the support of the NSF, STFC, INFN and CNRS for provision of computational resources.
Funding for this project was provided by the Charles E. Kaufman Foundation of The Pittsburgh Foundation and the Institute for Computational and Data Sciences at Penn State.
This article has been assigned the document number LIGO-P2100163-v8.

{\it We would like to thank all of the essential workers who put their health at risk during the COVID-19 pandemic, without whom we would not have been able to complete this work.}

\appendix
\section{Connecting $f_{\mathrm{PBH}}$ with Advanced LIGO and Advanced Virgo rate constraints}
\label{PBHAppendix}
We model an equal mass population of PBHs that are initially uniformly distributed in comoving volume. We parametrize the abundance of this population as a fraction of the total dark matter, i.e. $f_\mathrm{PBH} = \Omega_\mathrm{PBH} / \Omega_\mathrm{DM}$. We model the merger rates by considering two nearest neighbor, gravitationally bound black holes that are torqued by the next closest black hole. From these assumptions, we find the merger rate distribution~\cite{Ioka:1998nz,Sasaki:2016jop}

\begin{equation}
        \mathrm{d}P = \begin{cases}
         \dfrac{3 \, f_{\mathrm{PBH}}^{37/8}}{58} \left[f_{\mathrm{PBH}}^{-29/8} \left(\dfrac{t}{t_\mathrm{c}}\right)^{3/37} - \left(\dfrac{t}{t_\mathrm{c}}\right)^{3/8} \right] \dfrac{\mathrm{d}t}{t}, & \ t < t_\mathrm{c} \\[15pt]
        \dfrac{3 \, f_{\mathrm{PBH}}^{37/8}}{58} \left[ f_{\mathrm{PBH}}^{-29/8} \left(\dfrac{t}{t_\mathrm{c}}\right)^{-1/7} -  \left(\dfrac{t}{t_\mathrm{c}}\right)^{3/8} \right] \dfrac{\mathrm{d}t}{t}, & \ t \geq t_\mathrm{c}
    \end{cases}
\end{equation}
where $t_\mathrm{c}$ is a function of the mass of the compact objects and the fraction of the dark matter they comprise:
\begin{equation}t_\mathrm{c} = \frac{3}{170}\frac{c^5}{(Gm_\mathrm{PBH})^{5/3}} \frac{f_{\mathrm{PBH}}^7}{(1+z_{\text{eq}})^4} \left(\frac{8 \pi}{3 H_{0}^2 \Omega_{\text{DM}}}\right)^{4/3}.
\end{equation}
Here, $c$ is the speed of light, $G$ is the gravitational constant, $m_\mathrm{PBH}$ is the mass of the black holes in our equal mass population, $f_{\mathrm{PBH}}$ is the parametrized abundance from above, $z_\mathrm{eq}$ is the redshift at matter-radiation equality, $H_{0}$ is the Hubble constant, and $\Omega_{\text{DM}}$ is the dark matter density. We use the Planck “TT,TE,EE+lowP+lensing+ext” cosmology~\cite{Ade:2015xua} to evaluate $t_\mathrm{c}$.

The above equation, when evaluated at present day and multiplied by the number density of PBHs, provides a theoretical merger rate for PBHs:
\begin{equation} \mathcal{R}_\mathrm{PBH} = n_{\mathrm{PBH}}  \left. \frac{\mathrm{d}P}{\mathrm{d}t} \right|_{t=t_0} .\end{equation}
We equate our observed upper limit on the merger rate to the theoretical merger rate and invert at each value of $m_\mathrm{PBH}$ to obtain the constraint curve shown in Fig.~\ref{fig:upperlimits}.
This PBH model is discussed in further detail in the literature~\cite{Ioka:1998nz,Nakamura:1997sm,Sasaki:2016jop,Magee:2018opb}.

\section{Constraining dissipative dark matter using gravitational-wave searches for \SSM{} binaries}\label{App.DBHs}

We use a Bayesian approach to get the posterior probability of the fraction of dark matter in dark black holes, $f$, and the possible minimum mass of the DBH distribution, $M_{\rm min}$, using modelled rates for dark-matter BH mergers and estimated $\langle VT \rangle$ from searches for \SSM{} binary black holes. The 2D distribution for $\{f_{\rm DBH},~M_{\rm min}\}$ is obtained by marginalising over two additional parameters needed to characterize the binary distribution: the slope of the initial mass function, $b$, and a parameter $r=M_{\rm max}/M_{\rm min}$ that sets the mass range of the initial population. The 4D distribution is $P(f, \overline{\theta}=\{M_{\rm min}, b, r\} | \mathcal{R}_i, VT({\cal M}=m_i))$, which can be written in terms of the independent distributions for $f$ and the set $\overline{\theta}=\{M_{\rm min}, b, r\}$, as well as the likelihood $\mathcal{L}(f,\overline{\theta} ; {\mathcal{R}VT})$
\begin{equation}
P(f, \overline{\theta}| \mathcal{R}, VT) \propto P(f)P(\overline{\theta})\mathcal{L}(f,\overline{\theta} ; {\mathcal{R}VT}).
\end{equation}
The rates $\mathcal{R}_i$ are computed in pre-defined chirp mass bins within the range $\mathcal{M} \in [0.2M_\odot, 2.5 M_\odot]$ which is representative of the \SSM{} search, and depend on the model parameters $f$ and $\overline{\theta}$. The rates are modelled as:
\begin{equation}
\begin{split}
	\mathcal{R}_i (\mathcal{M} = m_i  | f, \overline{\theta}) =
	P_i (m_i | t_{\rm m}, \overline{\theta})
	\left(\frac{\mathrm{d}P(t_{\rm m} = 10~{\rm Gyr} | \overline{\theta})}{\mathrm{d}t}\right) \\
	\times
	\left(\frac{\rho_{\rm DM} \times f \times f_{\rm binary}}
	{\langle M \rangle}\right)
\end{split}
\end{equation}
where $\rho_{\rm DM} = 3.3\times 10^{19}\ M_\odot\ {\rm Gpc^{-3}}$ is the density of dark matter in the universe, and $f_{\rm binary} = 0.26$ is the number of dark black hole binaries divided by total DBHs. This is choice is informed by numerical studies of binary formation in Population III stars~\cite{Stacy:2012iz}. This number is of course uncertain, but other studies of Population III binaries (e.g., ~\cite{Belczynski:2016ieo}) often assume that binaries make about $1/3$ of all systems, which would correspond to the nearly identical $f_{\rm binary} = 0.25$. As $f_{\rm binary}$ is an overall multiplicative factor, the plotted constraint can be directly scaled for any other choice of $f_{\rm binary}$. The chirp mass distribution of binary systems that would merge within some merger time $t_{\rm m}$ is $P(\mathcal{M} | t_{\rm m}, \overline{\theta})$. Since these objects likely form between $20\lesssim z\lesssim30$, we may use $t_\mathrm{m}=10$ Gyr, roughly the age of the universe, and the exact formation time makes a negligible shift in this number. The probability that the merger time of the binary is 10 Gyr is denoted as $P(t_{\rm m} = 10\ {\rm Gyr} | \overline{\theta})$, and $\langle M \rangle$ is the mean component mass of dark-matter BHs given the initial mass distribution, given some $\overline{\theta}$. 

The $\langle VT \rangle$ estimated from \SSM{} searches for compact binary coalescences were weighted according to the allowed mass-ratios and their probabilities for a given population described by $\overline{\theta}$.
\begin{equation}\label{eq:VT2}
	VT_i(\mathcal{M}=m_i | \overline{\theta}) = \int^{q_{\rm max}}_1 \mathcal{P}(q|m_i,  t_{\rm m}, \overline{\theta})V T(m_i)\mathrm{d}q .
\end{equation}

We assume a Poisson distribution for event counts such that the rate posterior for zero detections becomes $\mathcal{P}(\mathcal{R} | VT) = VT \exp(-\mathcal{R}\times VT)$. The above definitions for $\mathcal{R}$ and $VT$ are used to compute the rate posterior in each bin. The likelihood of $f, \overline{\theta}$ is computed by taking a product of rate posteriors over all chirp mass bins.
\begin{equation}\label{eq:likelihood}
	\mathcal{L}(f,\overline{\theta} ; {\mathcal{R}VT})
	= \displaystyle\prod_{i}
	 \frac{\int^{\infty}_{\mathcal{R}_i}\mathcal{P}_i(\mathcal{R} | f, \overline{\theta}, VT_i) \mathrm{d}\mathcal{R}}
	 {\int_{0}^{\infty}\mathcal{P}_i(\mathcal{R} | f, \overline{\theta}, VT_i) \mathrm{d}\mathcal{R}} .
\end{equation}

We use $b \in [-1,2]$ for initial mass distribution of DBHs $\mathcal{P}(m) \propto m^{-b}$ and $r \in [2,1000]$ to constrain $f \in [10^{-10}, 1]$ and $M_{\rm min} \in [10^{-3} M_\odot, 3.1 M_\odot]$, shown in Fig~\ref{fig:dissipativeDM}. The range of initial mass function slopes $b$ is inclusive of all values found in the literature on Population III star binaries~\cite{Greif:2011it, Stacy:2012iz,Hartwig:2016nde}, while the range chosen for $r$ includes Population III star values~\cite{Hartwig:2016nde} and was shown in Ref.~\cite{Singh:2020wiq} to be sufficient so that results are not too sensitive to changes in the range.

\clearpage
\bibliography{references}

\end{document}